\documentstyle[aps,pre,epsfig]{revtex}

\newcommand{\be}{\begin{equation}}
\newcommand{\ee}{\end{equation}}
\newcommand{\lb}{\ell_B}

\newcommand{\Zeff}{Z_{\hbox{\scriptsize eff}}}
\newcommand{\Zbare}{Z_{\hbox{\scriptsize bare}}}
\newcommand{\Zsat}{Z_{\hbox{\scriptsize sat}}}

\begin{document}
\title{Effective interactions and phase behaviour for a model clay suspension 
in an electrolyte
\footnote{This paper is dedicated to Jean-Pierre Hansen on the occasion 
of his 60th birthday}
}
\author{E. Trizac$^1$, L. Bocquet$^2$, R. Agra$^1$, J-J. Weis$^1$ and M. Aubouy$^3$ }
\address{$^1$Laboratoire de Physique Th{\'e}orique, UMR CNRS 8627,
B{\^a}timent 210,Universit{\'e} Paris-Sud,91405 Orsay Cedex, France}
\address{$^{2}$ Laboratoire de Physique de l'Ens Lyon, UMR 
CNRS 5672, 46 All\'ee d'Italie, 69364 Lyon Cedex, France}
\address{$^{3}$ S.I.3M., D.R.F.M.C.,CEA-DSM Grenoble, 
17 rue des Martyrs, 38054 Grenoble Cedex 9, France}

\date{\today}
\maketitle

\begin{abstract}
Since the early observation of nematic phases of disc-like clay colloids 
by Langmuir in 1938,
the phase behaviour of such systems has resisted theoretical understanding. 
The main reason is that there is no satisfactory generalization for 
charged discs of the
isotropic DLVO potential describing the effective interactions between a pair
of spherical colloids in an electrolyte. In this contribution, we show how to
construct such a pair potential, incorporating 
approximately both the non-linear effects of
counter-ion condensation (charge renormalization) and the anisotropy of the
charged platelets. The consequences on the phase behaviour of Laponite dispersions
(thin discs of 30 nm diameter and 1 nm thickness) are discussed, 
and investigation into the mesostructure via Monte Carlo simulations
are presented.

\end{abstract}

\section{Introduction}
With the possibility to form an orientational ordering, the phase behaviour of
anisotropic colloids is richer than its counterpart for spherical particles. 
Surprisingly, the isotropic-nematic transition expected on purely entropic grounds 
(excluded volume effects) has been extensively reported for rod-like colloids 
in the last sixty years \cite{Bernal}, 
but could only be observed recently for discotic particles in apolar media
\cite{Kooij}. This experimental 
work directly points to the subtle effect of electrostatic interactions, given that
in the widely studied model system of colloidal platelets,
namely aqueous clay dispersions,
the isotropic/nematic phase separation is hindered by a ubiquitous ``fluid-solid''
transition. 

In spite of an important experimental and theoretical effort in the last 10 years,
triggered by the emergence of Laponite as a model system for disc-like colloidal
suspensions\cite{Lapo}, the above transition is far from being well 
understood
\cite{Mourchid1,Dijkstra1,Trizac0,Gabriel,Pignon,Dijkstra2,Hansen2,Trizac1,Mourchid2,%
Carvalho1,Bonn,Fossum,Saunders,Kutter,Levitz1,Nicolai1,Knaebel,Carvalho2,%
Harnau,Meyer,Nicolai2,Nicolai3,Bellon,Abou,Leger,Levitz2,Michot,Rowanbis}. 
In particular, the precise nature of the phases observed experimentally
is debated \cite{Mourchid1,Bonn,Nicolai2}, so that the relevance
of the traditional terminology ``sol-gel'' to describe the transition 
is somewhat controversial. 

On the other hand, theoretical approaches to
describe the phase behaviour of charged disc-like particles in an electrolyte
are in their infancy \cite{Dijkstra1,Dijkstra2,Kutter,Harnau}, essentially
because there is no satisfactory generalization for discs of the isotropic 
DLVO potential \cite{Belloni2} describing the effective interactions between 
a pair of spherical colloids. In the simplest situation of two coaxial 
and parallel uniformly charged platelets, the effective Coulomb force has been 
computed within linear \cite{Trizac1} and non-linear \cite{Carvalho2}
Poisson-Boltzmann (PB) theory. A screened electrostatic pair potential
has been worked out \cite{Rowan}, allowing to compute analytically the
interaction energy for discs of arbitrary orientation; this approach 
holds at the level of linearized PB theory (weak electrostatic coupling), 
and is expressed as a 
perturbative expansion
in the parameter $\kappa r_0$, $r_0$ being the disc radius and $\kappa$ the inverse
Debye length in the electrolyte. It consequently becomes less
reliable when $\kappa r_0$ becomes of order 1 or larger (the typical situation 
for Laponite solutions), or when the charge on the 
platelets becomes too large (high electrostatic coupling, also typical 
of clay systems). 
The latter limitation may be circumvented 
by introducing the concept of charge renormalization 
\cite{Alexander,Belloni1,Hansen1,Belloni2}, while the former requires the 
re-summation 
of all the powers of $\kappa r_0$ involved in the ``multi-polar'' expansion propounded 
in \cite{Rowan}. In this work, we discuss how to take both aspects into account
and investigate the consequences on the phase behaviour. 
Our goal is to understand the effect of colloid anisotropy on a possible 
phase transition. We show that considering
electrostatic effects only and discarding van der Waals interactions
yields qualitative differences between spherical and discotic colloids
and that the equivalent of the fluid-solid transition 
for spheres is nevertheless reminiscent of the 
``sol-gel'' transition reported in \cite{Mourchid1}.

The paper is organized as follows. In section \ref{sec:lpb}, we obtain
a pair potential for discs valid at large distances within linearized 
PB theory. 
We show that at asymptotically large separations, this potential
remains strongly anisotropic, unlike its counterpart in vacuum or 
in a plain dielectric medium (no salt). The approach is generalized 
in section \ref{sec:gener} to an arbitrary ``one-'' or``two-dimensional'' colloid 
(i.e. with vanishing internal volume). 
The non-linear phenomenon of counter-ion
``condensation'' leading to a renormalization of the bare platelet charge
is then considered in section \ref{sec:Zeff} and the resulting effective charge
successfully tested against ``exact'' numerical simulations reported
in the literature. In section \ref{sec:phase}, we propose a first and
simplified investigation into the phase behaviour by mapping the interaction
energy onto an isotropic Yukawa potential. 
Finally, the full anisotropic potential including charge renormalization
is used to investigate the mesostructure of an assembly of interacting 
discs by means of Monte Carlo computer simulations (section \ref{sec:mc}),
and conclusions are drawn in section \ref{sec:concl}.

\section{Large distance screened electrostatic potential}
\label{sec:lpb}

We consider first a unique platelet ${\cal P}$ 
of radius $r_0$ and surface charge $\sigma$
in a 1:1 electrolyte of bulk density $n$ (infinite dilution limit). The solvent 
is assumed to be a dielectric continuum of permittivity $\varepsilon$.
Within linearized Poisson-Boltzmann theory (LPB), the dimensionless 
electrostatic potential obeys the following Poisson's equation
\be
\nabla^2 \phi = \kappa^2 \phi,
\label{eq:lpb}
\ee
where $\kappa^2 = 8 \pi \lb n$ is the inverse Debye length squared and
$\lb = e^2 /(kT \varepsilon)$ denotes the Bjerrum length corresponding to the
distance where the bare Coulomb potential felt by two elementary point charges 
becomes equal to the thermal energy $kT$ ($\lb \simeq 7\,$\AA\ in water at room 
temperature). The potential is chosen to vanish at infinity and $e$ denotes 
the elementary charge. 
 
We are interested in the behaviour of $\phi$ at large distances,
which is relevant to describe the interactions in dilute suspensions. 
The solution of Eq. (\ref{eq:lpb}) has been worked out in the form 
of an integral representation in \cite{Hsu,Trizac1}, or as a
multi-polar-like expansion \cite{Rowan}. It is however more convenient
to obtain an explicit expression, that is a good starting point
to derive the pair interaction. This can be achieved by writing the
solution as a convolution of the surface charge of the platelets with the
screened Coulomb potential:
\be
\phi({\bf r}) \,=\, \lb \int_{\cal P} \frac{\sigma}{e} \,
\frac{\exp(-\kappa|{\bf r}-{\bf s}|)}{|{\bf r}-{\bf s}|} \,d^2{\bf s}.
\label{eq:convol}
\ee
It is important to realize that such a convolution procedure would
give incorrect results for polyions of non vanishing excluded volume
(e.g. spheres or rods of non zero radii, see the appendix).
Introducing the unit vector ${\bf \widehat n}$ normal to the disc surface,
and the angle $\theta \in [0,\pi/2]$ between the corresponding direction and 
the position vector ${\bf r}$ with origin at platelet centre, 
we obtain the leading contribution
\be
\phi({\bf r}) \, \stackrel{\kappa r \gg 1}{\sim}\, \lb\,
\Zbare\, 2\,\frac{\hbox{I}_1(\kappa r_0 \sin \theta)}{\kappa r_0 \sin \theta}\,
\frac{e^{-\kappa r}}{r} \qquad \hbox{with} \qquad \Zbare e = \pi r_0^2 \sigma,
\label{eq:lpbdisc}
\ee
where $\hbox{I}_1$ denotes the modified Bessel function of the first kind,
such that $\hbox{I}_1(x) \sim x/2$ for $|x| \ll 1$.
A peculiarity of screened Coulomb potential appears at this point: 
whereas the large $r$ potential becomes isotropic and 
behaves as $\Zbare /r$ in vacuum
(situation corresponding to the limit $\kappa \to 0$), 
the anisotropy is present at all distances in an electrolyte
($\kappa \neq 0$): the $\theta$ and $r$ dependence factorize in equation 
(\ref{eq:lpbdisc}). 
In practice, the anisotropy of the potential
is generically significant when the size of the object under consideration
is larger than the Debye length (i.e. $\kappa r_0 >1$ here). 
For Laponite discs of radius 150$\,$\AA\, this crossover
corresponds to an ionic strength $I^*=10^{-4}\,$M. As noted in
\cite{Levitz1}, $I^*$ corresponds experimentally 
to a threshold value delimiting
qualitatively different phase behaviours.
At fixed distance $r$, the potential (\ref{eq:lpbdisc}) is minimum
for $\theta=0$ which corresponds to the configuration where the
``average'' distance between the point where $\phi$ is computed and
the platelet is maximal; on the other hand, the potential is maximum
for $\theta=\pi/2$. From (\ref{eq:lpbdisc}) the ratio between these 
extremal values is $r$-independent, and reads $2\hbox{I}_1(\kappa r_0)/(\kappa r_0)$
which can be as large as 10 for $\kappa r_0 =5$ (important anisotropy). 

\begin{center}
\begin{figure}
\input{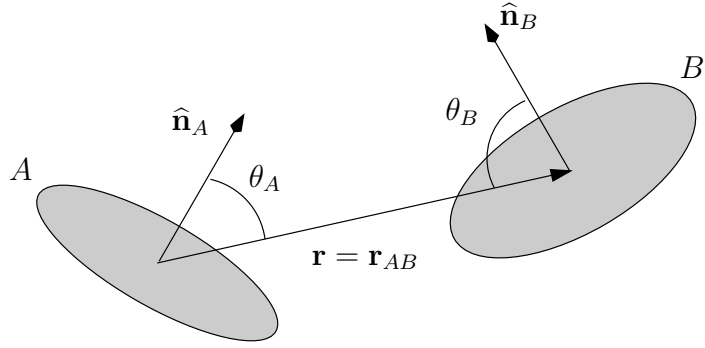}
\vspace{5mm}
\caption{Definition of the coordinates used in the two body problem
}
\label{fig:2plates}
\end{figure}
\end{center}

This result may be used to compute the potential energy of interactions between
two platelets $A$ and $B$ with arbitrary relative orientations, as shown in 
Fig. \ref{fig:2plates}. This energy is obtained by integrating the screened
potential created by disc $A$ defined by Eq. (\ref{eq:lpbdisc}) 
over the surface charge distribution of the
second disc ($B$):
\be
V_{AB}(r,\theta_A,\theta_B) \,=\, \int_{B} \frac{\sigma}{e} 
\phi_A({\bf r}+{\bf s}) \, d^2{\bf s},
\label{eq:vabconvol}
\ee
where $V_{AB}$ is the dimensionless potential expressed in $kT$ units.
Again, such a procedure would not give the proper interactions in the
case of polyions with internal volume. At large distances, we obtain
\be
V_{AB}(r,\theta_A,\theta_B) \,\sim\, 
\Zbare^2\, \lb\, 4\,\frac{\hbox{I}_1(\kappa r_0 \sin \theta_A)}{\kappa r_0 \sin \theta_A}\,
\frac{\hbox{I}_1(\kappa r_0 \sin \theta_B)}{\kappa r_0 \sin \theta_B}\,
\frac{e^{-\kappa r}}{r}.
\label{eq:vab}
\ee
Clearly, the relative orientation of the platelets is not
completely specified by the three parameters $r$, $\theta_A$ and $\theta_B$,
but the omitted Euler angles only appear in higher order terms
such as $\exp(-\kappa r)/r^2$, $\exp(-\kappa r)/r^3$\ldots
In particular the energy (\ref{eq:vab}) appears insensitive to 
a precession of the discs around their centre to centre direction.

\begin{center}
\begin{figure}
\input{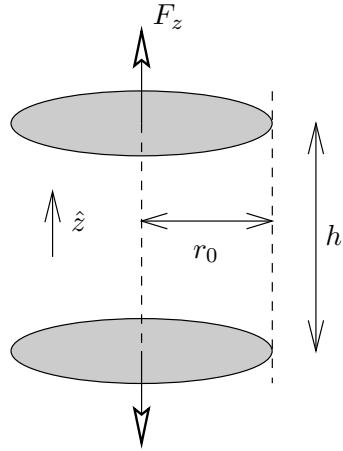}
\caption{Geometry used to compute the force between two discs in section 
\ref{sec:Zeff}.}
\label{fig:arg-2discs}
\end{figure}
\end{center}

At fixed centre to centre distance, the above energy (always repulsive 
\cite{Trizac3}) is maximized for co-planar discs
($\theta_A = \theta_B = \pi/2$, which corresponds to the maximum overlap
of electric double layers) and minimized when the discs are co-axial
and parallel ($\theta_A=\theta_B=0$, see the configuration represented
in Fig. \ref{fig:arg-2discs}). A situation of intermediate 
electrostatic energy is that of {\sf T}-shape perpendicular discs
($\theta_A=0$ and $\theta_B=\pi/2$). These results are consistent 
with the numerical linearized Poisson Boltzmann pair potential 
reported in \cite{Hsu}, where it was also shown that the interactions
between platelets at constant surface charge or constant potential
were qualitatively very similar.

In reference \cite{Rowan}, the potential energy $V_{AB}$ was obtained in the 
form of a multi-polar expansion. Unlike its unscreened counterpart where
the multipole of order $l$ has a $1/r^{l+1}$ large distance contribution 
to the potential, this expansion 
is such that the multipole of order $l$ contributes to orders 
$\exp(-\kappa r)/r, \exp(-\kappa r)/r^2\ldots\ \exp(-\kappa r)/r^{l+1}$. 
The large $r$ potential can in principle be obtained by 
re-summation of all multi-polar contributions, which should lead
to expression (\ref{eq:vab}). Alternatively, truncating the
multi-polar expansion at a given order amounts to expanding
(\ref{eq:vab}) in powers of $\kappa r_0$. Including monopole-monopole,
monopole-quadrupole and quadrupole-quadrupole interactions, 
the following expression was 
obtained in \cite{Rowan}:
\be
V_{AB}(r,\theta_A,\theta_B) \,=\, \Zbare^2\, \lb\, \left[
1  + \frac{\kappa^2 r_0^2}{8}\, (\sin^2\theta_A + \sin^2\theta_B)
+ \frac{\kappa^4 r_0^4}{64}\, (  \sin^2\theta_A \sin^2\theta_B)
\right] \,\frac{e^{-\kappa r}}{r}.
\label{eq:vabrowan}
\ee
From (\ref{eq:vab}), we obtain
\be
V_{AB}(r,\theta_A,\theta_B) \,=\, \Zbare^2\, \lb\, \left[
1+ \frac{\kappa^2 r_0^2}{8}\, (\sin^2\theta_A + \sin^2\theta_B)
+ \frac{\kappa^4 r_0^4}{64}\, (  \sin^2\theta_A \sin^2\theta_B +
\frac{1}{3}\sin^4\theta_A \sin^4\theta_B) + {\cal O}(\kappa^6 r_0^6)
\right] \,\frac{e^{-\kappa r}}{r} 
\ee
Both expressions agree at order $(\kappa r_0)^2$, and the difference
at order $(\kappa r_0)^4$ is the monopole-hexadecapole
contribution which has not been included in (\ref{eq:vabrowan}). 
This comparison illustrates the perturbative nature of the potential
derived in \cite{Rowan}, and the fact that the corresponding multipolar
contributions are implicitly re-summed in expression (\ref{eq:vab}).

\section{Generalization to a polyion of arbitrary shape}
\label{sec:gener}

The method used in the previous section may be generalized to find the
far field electrostatic potential of an arbitrary polyion with vanishing 
internal volume and bare charge 
$Z=\int_{\hbox{\scriptsize polyion}} \sigma({\bf s}) \,d^2{\bf s} $.
In general,  
the convolution solution (\ref{eq:convol}) admits
the large distance behaviour:
\be
\phi({\bf r}) \,\sim \, Z\, \lb\, f({\bf \widehat r},\kappa) \, \frac{e^{-\kappa r}}{r}
\,+\, {\cal O}\left(\frac{e^{-\kappa r}}{r^2}\right) \quad \hbox{for}
\quad \kappa r \gg 1.
\ee
In this expression, the anisotropic part of the potential again factorizes 
from the $r$-dependence and is given by
\be
f({\bf \widehat r},\kappa) = \int_{\hbox{\scriptsize polyion}} 
\frac{\sigma({\bf s})}{Z e}\, \exp\left(-\kappa\, {\bf \hat r}\cdot{\bf s}\right)
\, d^2{\bf s},
\ee
where ${\bf \hat r}$ denotes a unit vector in the direction 
of the position ${\bf r}$ where the potential is computed,
and $\sigma({\bf s})$ is the surface charge density at point
${\bf s}$ on the colloid. As expected, the isotropic bare Coulomb potential
is recovered in the limit of low electrolyte density:
\be
\lim_{\kappa \to 0} \, f({\bf \widehat r},\kappa) \,=\, 1,
\ee
with $Z$ the bare charge. 

For a circle of radius $r_0$ and uniform line charge we find
\be
f(\theta,\kappa r_0) \,=\, \hbox{I}_0(\kappa r_0 \sin\theta),
\ee
where $\theta$ is the angle between ${\bf r}$ and the normal to the plane
containing the circle. Finally, for a uniformly charged rod of vanishing radius,
and length $2l$:
\be
f(\theta,\kappa l) \,=\, \frac{\sinh(\kappa l \cos\theta)}{\kappa l \cos\theta},
\ee
where $\theta$ is now the angle between ${\bf r}$ and the rod direction.

\section{Towards charge renormalization for platelets}
\label{sec:Zeff}

The bare charge of Laponite platelets can be considered to be of the
order of $\Zbare =700$ to 1000 negative elementary charges \cite{Meyer,Nicolai2}.
At the level of PB approximation, this corresponds to a high electrostatic
coupling ($\phi$ larger than unity \cite{Carvalho1,Carvalho2}), 
where the linearization procedure
underlying Eq. (\ref{eq:lpb}) fails. However, a few Debye lengths away 
from the colloid, the potential has sufficiently decreased so that (\ref{eq:lpb})
is recovered and the one-body potential takes the form
$$
\phi(r,\theta) \,=\, \Zeff \,f(\theta,\kappa r_0)\, \frac{e^{-\kappa r}}{r}.
$$
The effective charge $\Zeff$ is (in absolute value) smaller than the bare
one \cite{Alexander,Belloni1,Hansen1}, and within PB theory, saturates
to a value independent of $\Zbare$ if the latter quantity is large enough. 
The contribution embodied in $f(\theta,\kappa r_0)$ results from the anisotropy
of the colloid, and is a priori an unknown function. In the limit of small
electrostatic coupling (small $\Zbare$) where $\Zeff/\Zbare \to 1$ by definition, 
the results of section \ref{sec:lpb} show that 
$f(\theta,X) = 2 \hbox{I}_1(X \sin\theta)/(X\sin\theta)$.
On the other hand, at higher $\Zbare$ corresponding to the saturation 
regime, the functional form of $f(\theta,X)$ is the signature of the
effective charge distribution on the platelet and related to possible
differences in counter-ion condensation around the centre of the discs
or in the vicinity of the edges. 

It has been shown within PB theory that in the colloidal limit $\kappa r_0 >1$,
highly charged colloids behave as far as their far field is concerned 
as constant potential particles whose value is close to $4 kT/e$ (i.e. 
$\phi=4$), irrespective of shape (planar, cylindrical, spherical) \cite{Trizac2}.
This prescription may be applied to the present case, with the restriction that
to our knowledge, the linearized Poisson-Boltzmann problem cannot be solved
analytically with the boundary condition of constant surface potential. 
However, the potential associated with the constant surface charge boundary
condition provides a reasonable first approximation providing the correct
qualitative variation of the effective charge with physico-chemical
parameters. 
The corresponding anisotropic contribution to the
large distance field is that computed in section \ref{sec:lpb} and the
calculation of 
$\Zeff$ follows from the knowledge of $\phi(r=0)$. From the 
analytical expression reported in \cite{Trizac1}, we obtain 
the saturation value of $\Zeff$
\be
\Zeff^{\hbox{\scriptsize saturation}}\,=\,\Zsat \,=\, 
\frac{r_0}{\lb}\, \frac{2 \kappa r_0}{1-\exp(-\kappa r_0)}.
\label{eq:Zsat}
\ee
It is noteworthy that for a typical salt concentration corresponding to $\kappa r_0=1$,
$\Zsat \simeq 100$, which is an order of magnitude smaller than the bare charge.
This asymmetry guarantees that the effective charge is in the
saturation regimes where it no longer depends on the value of $\Zbare$
\cite{Alexander,Trizac2}
(the precise experimental determination of $\Zbare$ is therefore
not required). This saturation picture is expected to be reliable 
in a 1:1 electrolyte (PB mean-field approximation generally fails
in electrolytes of higher valence), when there is a clear scale
separation between the Bjerrum length and the size of the charged
object (this constraint being fulfilled for Laponite) \cite{Groot}.

\begin{center}
\begin{figure}[h]
\epsfig{figure=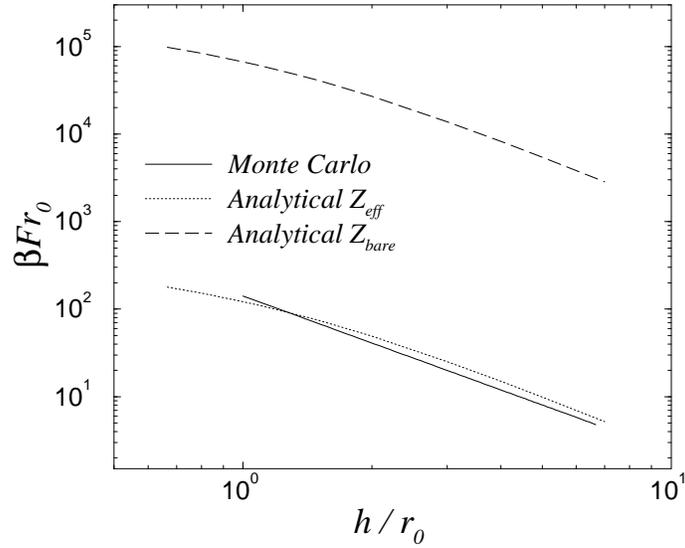,width=9cm,angle=0}
\caption{Comparison of the effective force between two parallel and coaxial
platelets obtained by ($N,V,T$) Monte Carlo simulations in \protect\cite{Meyer},
with the general expression for the LPB force supplemented with the
effective charge given by (\ref{eq:Zsat}). $h$ is the distance between 
the two plates (see Fig. \ref{fig:arg-2discs}). The same results are shown
neglecting charge renormalization, which amounts to considering $\Zeff=\Zbare$
(upper dashed curve).}
\label{fig:force}
\end{figure}
\end{center}

The above prediction may be checked by inserting $\Zsat$ into the analytical 
expression giving the LPB force between two parallel platelets \cite{Trizac1}
(configuration depicted in Fig. \ref{fig:arg-2discs}):
\be
\frac{r_0 F_z}{kT}\, = \,\frac{4 \pi \lb}{r_0}\, \Zsat^2 \,
\int_{0}^{\infty} J_1^2(x) \,\frac{1}{x} \,\exp\left\{-(h/r_0)
\sqrt{x^2+\kappa^2 r_0^2} \right\}\, dx
\ee
and comparing the results with the Monte Carlo simulations of
Meyer {\it et al} \cite{Meyer}, who considered the situation
of vanishing salt. In this limit, where the quality of our prescription
for $\Zeff$ is expected to deteriorate, we get from (\ref{eq:Zsat})
$\Zsat \to 2 r_0 /\lb \simeq 42$. The comparison is displayed 
in Fig. \ref{fig:force} which shows a good agreement.
Neglect of charge renormalization effects lead to an overestimation 
of the force by more than two orders of magnitude [more precisely,
by a factor $(\Zbare/\Zsat)^2$, see the difference between dashed and dotted
curves in Fig \ref{fig:force}], which points to the prime importance 
of such a phenomenon.

\section{Tentative investigation into the phase behaviour}
\label{sec:phase}

\begin{center}
\begin{figure}
\input{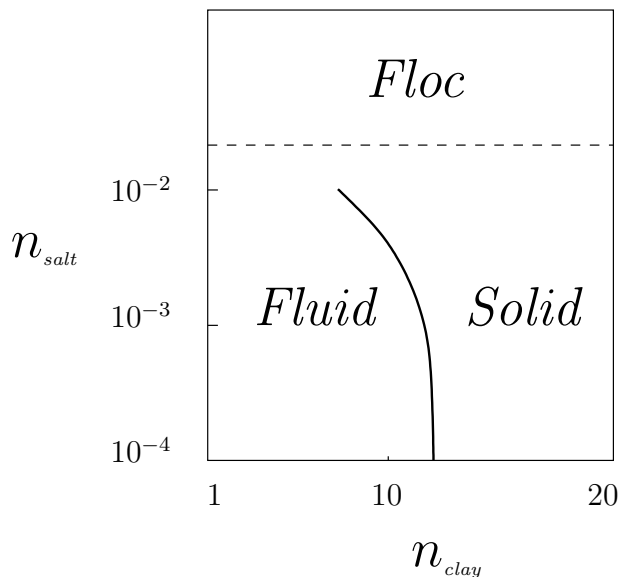}
\caption{
Schematic phase diagram of Laponite suspensions, reproduced from 
Ref. \protect\cite{Mourchid1}, where the fluid-solid transition was 
referred to as a sol-gel transition. The salt concentration on the $y$-axis is
expressed in mol (dm)$^{-3}$, and the scale for Laponite  concentrations 
on the $x$-axis is in g.l$^{-1}$. The solid under consideration is isotropic,
but at higher clay densities, a nematic solid may be formed \protect\cite{Gabriel}.
The corresponding diagram for spherical colloids would exhibit a Fluid/Solid line
of opposite (positive) slope \protect\cite{Monovoukas}.
}
\label{fig:phases}
\end{figure}
\end{center}

At this point, it is interesting to investigate at least qualitatively
the difference in the phase behaviour between spherical and disc-like
charge stabilized colloids. For spheres, the fluid-solid transition driven
by repulsive electrostatic interactions favors the isotropic fluid upon addition 
of salt in the solution \cite{Monovoukas}. The opposite is observed for Laponite
dispersions, where an increase of the ionic strength lowers the density where the
``solid'' phase appears \cite{Mourchid1,Mourchid2}. Given that van der Waals
interactions are believed to be irrelevant in the corresponding parameter 
range \cite{Mourchid2}, this effect
(illustrated in Fig. \ref{fig:phases}), appears at first to contradict 
standard DLVO phenomenology, where a screening of electrostatic repulsion 
(decrease of Debye length $\kappa^{-1}$) is 
expected to promote the simple fluid phase (sol), as happens for spheres. 
However, the effective charge of an arbitrary charged object is generically
a growing function of salt concentration [see for instance expression
(\ref{eq:Zsat}), which increases with $\kappa$], as a result of an 
enhanced screening of colloid/micro-ions
attraction which diminishes the amount of counter-ion ``condensation''. 
This increase of the effective charge favours the solid phase and is 
thus antagonistic to the abovementioned decrease of the range of electrostatic
repulsion driven by the decrease of $\kappa^{-1}$. 
The competition between the increase of the amplitude of repulsion 
and the decrease of its range upon adding salt
is a possible scenario to interpret the phase behaviour
of discs, within the DLVO picture and without any attractive interactions.
More quantitative results are given below.

In the following analysis, we treat spheres and discs on equal footings
(both are considered to have the same radius $r_0$). 
From the above analysis, we consider the following pair potential 
for discs:
\be
V_{AB}(r,\theta_A,\theta_B) \, = \, \Zsat^2
\, \lb\, 4\,\frac{\hbox{I}_1(\kappa r_0 \sin \theta_A)}{\kappa r_0 \sin \theta_A}\,
\frac{\hbox{I}_1(\kappa r_0 \sin \theta_B)}{\kappa r_0 \sin \theta_B}\,
\frac{e^{-\kappa r}}{r},
\label{eq:dlvodiscs}
\ee
where $\Zsat$ is given by Eq. (\ref{eq:Zsat}); its counterpart for charged spheres
is the standard isotropic DLVO expression \cite{Hansen1,Belloni2}
\be
V_{AB}(r) \, = \, \Zsat^2 \, \lb\,
\, \left(\frac{e^{\kappa r_0}}{1+\kappa r_0}\right)^2\,
\frac{e^{-\kappa r}}{r},
\label{eq:dlvo}
\ee
where the saturation value of the effective charge has been derived analytically
in \cite{Trizac2} as a function of salt concentration 
[we again consider colloids with a high bare charge such that
$\Zeff$ coincides with its saturation value].

\begin{center}
\begin{figure}[h]
\epsfig{figure=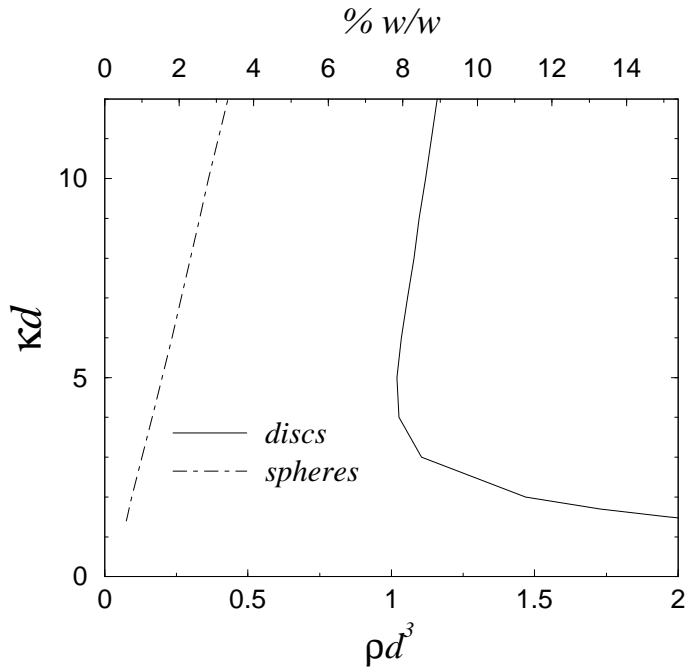,width=9cm,angle=0}
\caption{Tentative phase diagram of highly charged colloids in a 1:1 electrolyte,
as obtained for particles interacting through a Yukawa potential in
\protect\cite{Robbins,Meijer,Bitzer}.
The equivalent of the fluid-solid transition 
for spheres is shown for discs [i.e. the underlying Yukawa potential is
given by (\ref{eq:dlvo}) for spheres and by (\ref{eq:average}) for discs].
Discs and spheres have the same radius $r_0=150\,$\AA\ (reasonable for Laponite
clays), and $d=2r_0$ 
denotes the diameter. The bottom $x$-label corresponds the dimensionless
number density, while the upper $x$-scale converts this quantity in terms
of the ratio of clay mass over solvent mass for Laponite platelets
(this scale is consequently irrelevant for spherical colloids). 
In both cases,
the line shown is the melting curve delimiting a solid at high density
and a fluid for more dilute suspensions.}
\label{fig:discsphere}
\end{figure}
\end{center}

Given that the phase behaviour of
particles interacting through Yukawa-like potentials has been extensively
explored by computer simulations \cite{Robbins,Meijer,Bitzer}, 
we can readily obtain the melting density corresponding to a 
given salt concentration for the potential (\ref{eq:dlvo}). 
The results (corresponding to spherical colloids) 
are shown in Fig. \ref{fig:discsphere}.
A related procedure may be used to estimate qualitatively the position of the
melting or freezing line for discs, by assuming that at low clay density
(typically where the fluid/solid transition takes place for Laponite
suspensions),
the rotational motion of the platelets occurs on a shorter time scale 
than the translational one, so that these objects feel an effective 
potential resulting from the angular average of the expression given in
Eq. (\ref{eq:dlvodiscs}), namely 
\begin{eqnarray}
V_{AB}^{\hbox{\scriptsize average}}(r) &=&
\left\langle V_{AB}(r,\theta_A,\theta_B)   \right\rangle_{\theta_A,\theta_B} \,=\,
\frac{1}{4}\,\int_{0}^\pi \sin\theta_A d\theta_A 
\int_{0}^\pi \sin\theta_B d\theta_B\, V_{AB}(r,\theta_A,\theta_B)  \nonumber\\
&=&   \Zsat^2\, \lb\, 4\left(\frac{\cosh(\kappa r_0)-1}{\kappa^2 r_0^2} 
\right)^2 \,\frac{e^{-\kappa r}}{r}.
\label{eq:average}
\end{eqnarray}
In doing so, we obtain an isotropic Yukawa potential where the energy scale 
--the term in parenthesis in (\ref{eq:average})--
reflects the original anisotropy of the pair potential. In the standard 
DLVO potential (\ref{eq:dlvo}) for spheres, the energy scale 
[the term $\exp(\kappa r_0)/(1+\kappa r_0)$] also depends on 
Debye length, but has a different physical origin and results from  
the exclusion of micro-ions from the interior of the spheres. 
Making use of the numerical Yukawa phase diagram \cite{Robbins,Meijer,Bitzer},
we obtain the melting line represented in Fig. \ref{fig:discsphere}
for the averaged potential (\ref{eq:average}), where the threshold density
decreases when the salt concentration (or equivalently $\kappa$) increases,
at least for $\kappa d <4$. At the same level of description, spherical
colloids show the opposite behaviour, for all values of $\kappa d$ 
(we also emphasize that for the spherical colloids used in the experiments
of Monovoukas and Gast \cite{Monovoukas}, 
the equivalent of the melting line reported in 
Fig. \ref{fig:discsphere} is in excellent agreement with its 
experimental counterpart \cite{Trizac2}). 

For both spheres and discs, the prefactor of the Yukawa
term $\exp(-\kappa r)/r$ is an increasing
function of $\kappa$ [see Eqs. (\ref{eq:dlvo}) and (\ref{eq:average})]. 
The subtle interplay between this increase and the decrease of
Debye length is nevertheless able to produce a ``re-entrant'' melting
line for discs only. In the limit of low $\kappa r_0$, both expressions
(\ref{eq:dlvo}) and (\ref{eq:average}) become $\lb \Zsat^2 \exp(-\kappa r)/r$;
the melting lines for spheres and discs however do not merge in this
limit in Fig. \ref{fig:discsphere}, due to the difference in the
limiting values of $\Zsat$ for both geometries. 

This approach consisting in averaging the two-body platelet
potential over angular degrees of freedom predicts a qualitative
change in the phase behaviour of platelet systems for a reduced 
density $\rho^* = \rho d^3$ of order 1, which corresponds to a clay 
mass fraction of 8\% for Laponite with diameter $d=300\,$\AA\ , 
see the upper $x$-label of Fig. \ref{fig:discsphere}.
This density is approximately 4 times
higher than the maximum density delimiting the fluid and solid regions
in the phase diagram of Laponite suspensions (Fig. \ref{fig:phases} and
references \cite{Mourchid1,Mourchid2}), so that the present approach does
not allow a quantitative comparison with experiments. It is however 
noteworthy that a density $\rho^* \simeq 1$ is much smaller than the 
isotropic/nematic coexistence density for uncharged plates 
($\rho^* \simeq 4$, \cite{Eppenga,Veerman}). At this stage,
it is impossible to be more specific concerning the nature of
the ``solid'' phase supplementing the fluid one at high densities.
This question will be addressed in the following section by 
computer simulations.

\section{Mesostructure: Monte Carlo simulations}
\label{sec:mc}

We have implemented standard
Monte Carlo simulations with typically $N=500$ platelets interacting through the
potential (\ref{eq:dlvodiscs})
where $\Zsat$ is given by Eq. (\ref{eq:Zsat}). This allows to test the 
validity of the approach proposed in section \ref{sec:phase}, where the
initial anisotropic potential was mapped onto the isotropic Yukawa
function (\ref{eq:average}). A typical run consisted in
$10^5$ cycles (both random displacement and rotation of the $N$ particles).  

For different values of salt concentration and platelet density, two diagnostics 
were used to characterize the simulated samples. First, the centre-to-centre
pair distribution function $g(r)$ was computed. Second, the orientational ordering 
was quantified through the statistical average of (twice) 
the second Legendre polynomial
$P(\psi) = 3 \cos^2\psi-1$ at a given centre to centre distance $r$,
where $\psi$ is the angle formed by the normals to the two discs.
The orientational pair correlation function denoted $g_{or}(r)$ 
follows from averaging over all pairs of platelets. 

\begin{center}
\begin{figure}[h]
\epsfig{figure=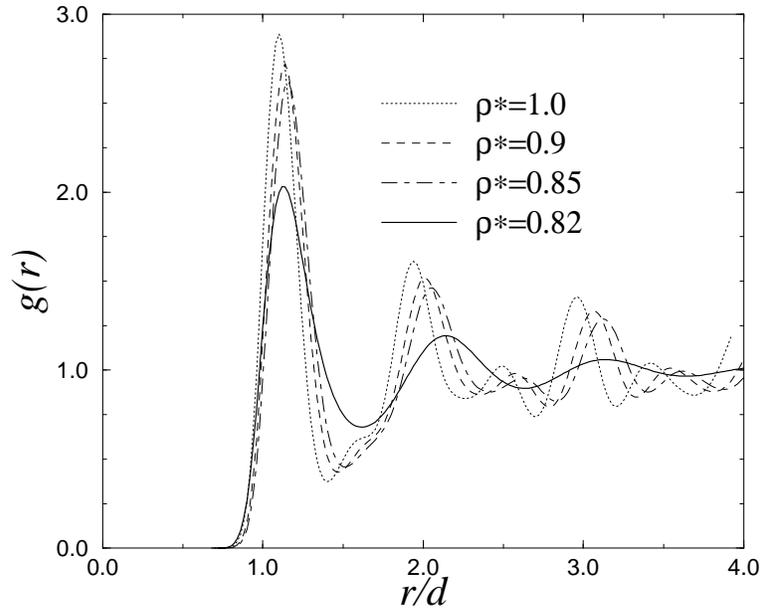,width=10cm,angle=0}
\caption{Center to center pair correlation function as a function 
of distance for $\kappa d =1$.}
\label{fig:jjwkd1}
\end{figure}
\end{center}

\begin{center}
\begin{figure}[h]
\epsfig{figure=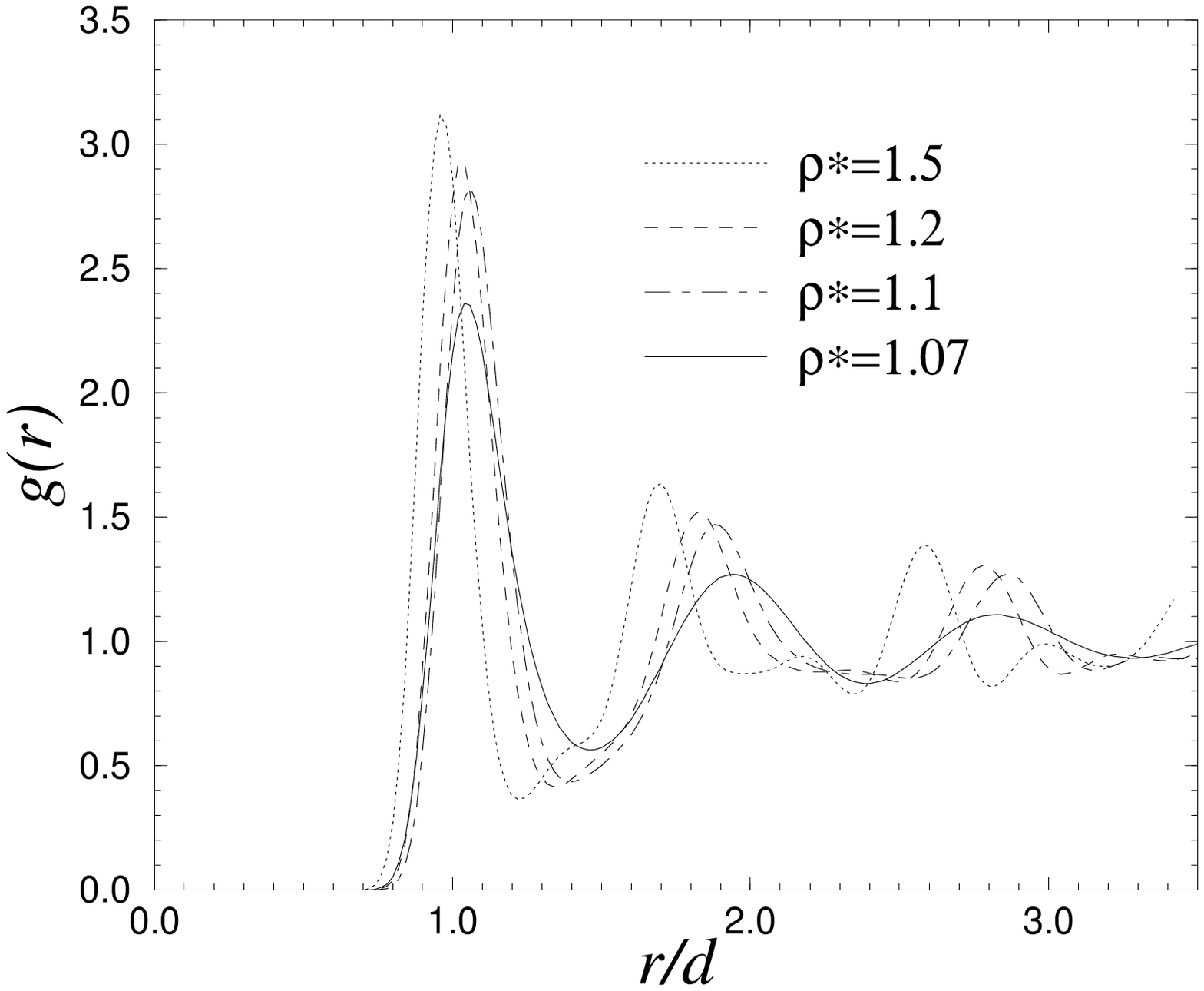,width=8.5cm,angle=0}
\epsfig{figure=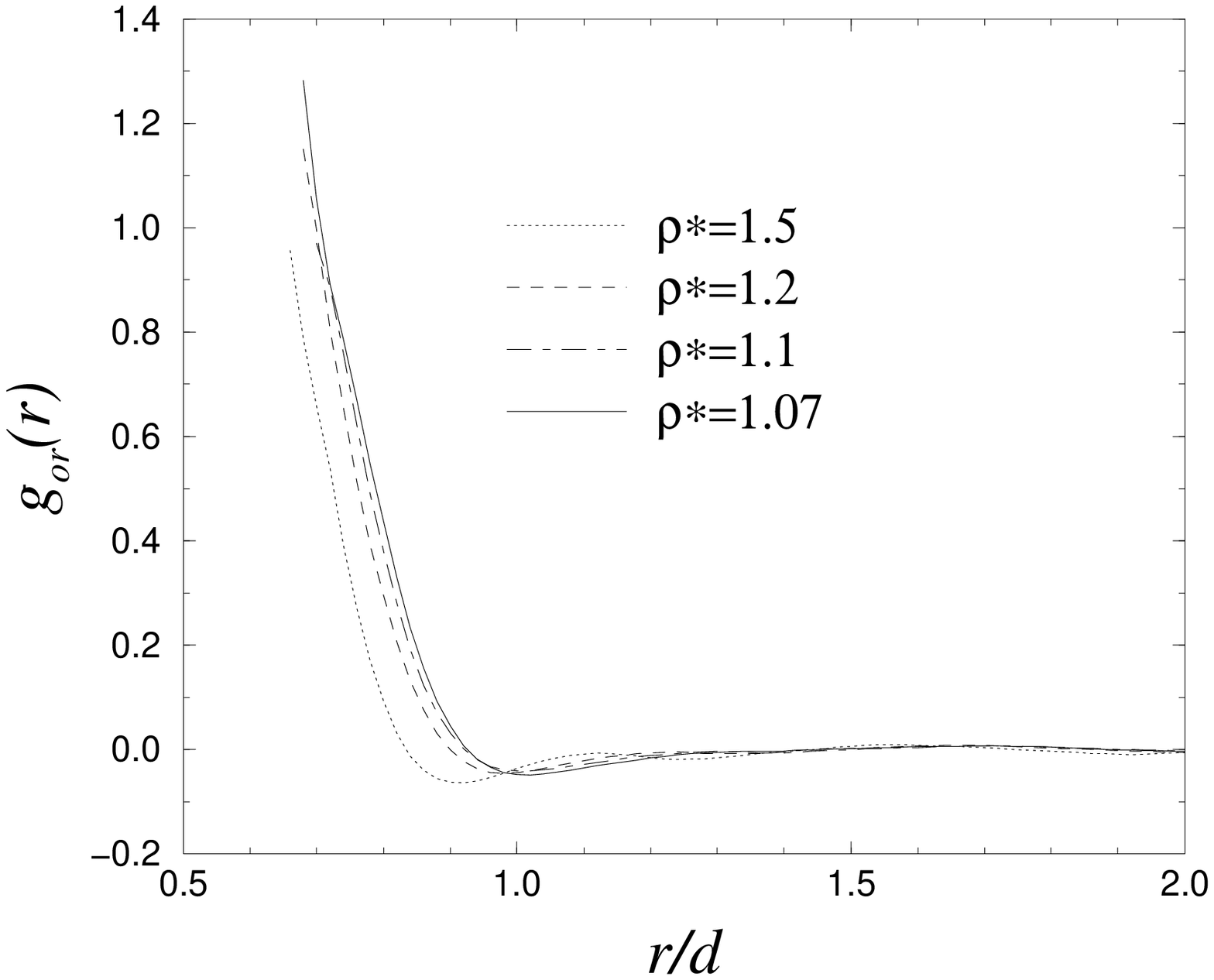,width=8.5cm,angle=0}
\caption{Same as Fig. \ref{fig:jjwkd1} for a higher 
salt concentration ($\kappa d =3$). The structure-less orientational distribution
is shown for completeness on the right graph.}
\label{fig:jjwkd3}
\end{figure}
\end{center}

The corresponding information on the mesostructure is displayed in Figs.
\ref{fig:jjwkd1} and \ref{fig:jjwkd3} for different densities and
two values of $\kappa$. A striking observation is that a slight increase 
of the density induces an important increase of the structure
(see the differences between the curves at $\rho^*=0.82$ and $\rho^*=0.85$
in Fig. \ref{fig:jjwkd1}; the same is seen at $\kappa d = 3 $ in Fig.
\ref{fig:jjwkd3} when the reduced density changes from 1.07 to 1.10). Such values 
of $\rho^*$ lie close to the threshold $\rho^*=1$ estimated in section
\ref{sec:phase}. The associated orientational distributions (not shown)
are structure-less for the densities considered here. 
These results indicate a qualitative change in the
structure of the fluid phase upon increasing the density, 
but it is still difficult to characterize the new ``phase'' emerging
without a thermodynamical study.

\begin{center}
\begin{figure}[h]
\epsfig{figure=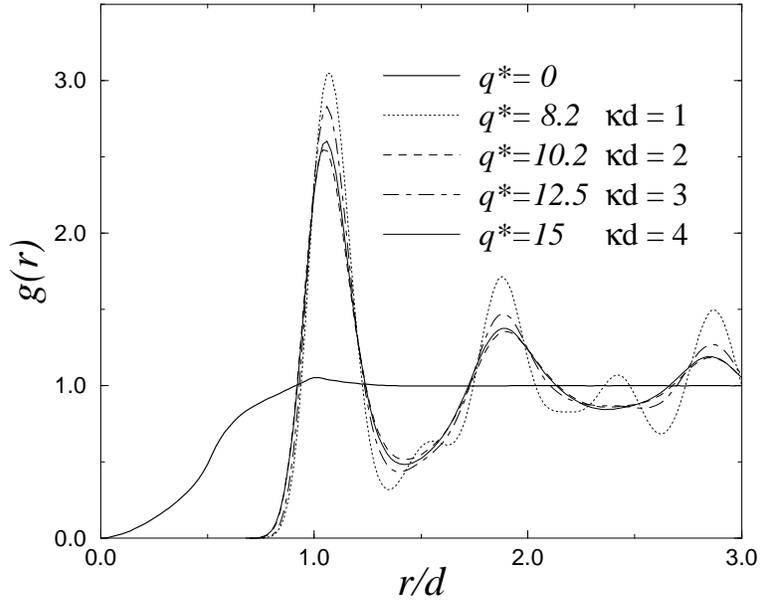,width=10cm,angle=0}
\caption{Pair distribution function for $\rho^* =1.1$. The reduced
charges defined as $q^*= \Zsat \sqrt{\lb/d}$ have been computed 
from (\ref{eq:Zsat}) and are given for completeness for every $\kappa$.
Also shown is the $g(r)$ for an uncharged platelet system 
at the same density (only excluded volume effects).  }
\label{fig:jjwrho1.1}
\end{figure}
\end{center}

We have tested the relevance of the scenario propounded at the beginning of
section \ref{sec:phase} (competition between an increase of the amplitude 
and a decrease of the range of the pair potential when $\kappa$ 
--or equivalently $n_{\hbox{\scriptsize salt}}$-- increases).
For a given density ($\rho^*=1.1$), 
the pair correlation functions are monitored for various 
values of $\kappa$. The corresponding charges change with $\kappa$, according
to Eq. (\ref{eq:Zsat}). Figure \ref{fig:jjwrho1.1} shows an interesting
feature. As expected, the 
structure is most pronounced for the lowest $\kappa$,
and decreases when $\kappa$ increases (see the difference between the
curves for $\kappa d=1$ and $\kappa d=2$). However, when salt concentration 
is further raised to $\kappa d=3$, the maximum of $g(r)$ and the whole structure
are enhanced, before decreasing again when $\kappa d=4$. 
The effect evidenced in Fig. \ref{fig:jjwrho1.1} is reminiscent of
the ``anomalous'' slope of the phase diagram reproduced in 
Fig. \ref{fig:phases}, and in qualitative agreement with the 
simplified phase diagram drawn in Fig. \ref{fig:discsphere}.

\begin{center}
\begin{figure}[h]
\epsfig{figure=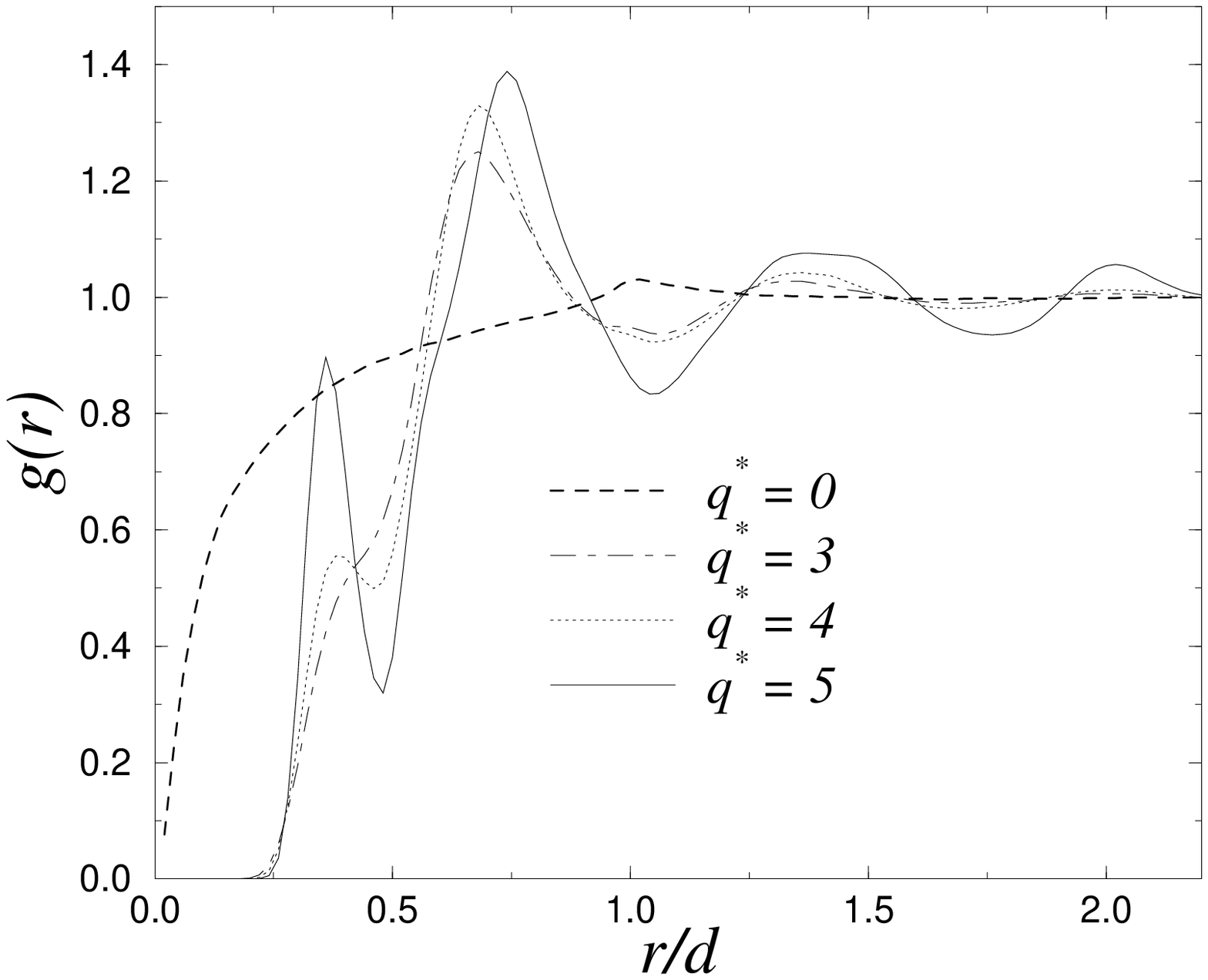,width=8.5cm,angle=0}
\epsfig{figure=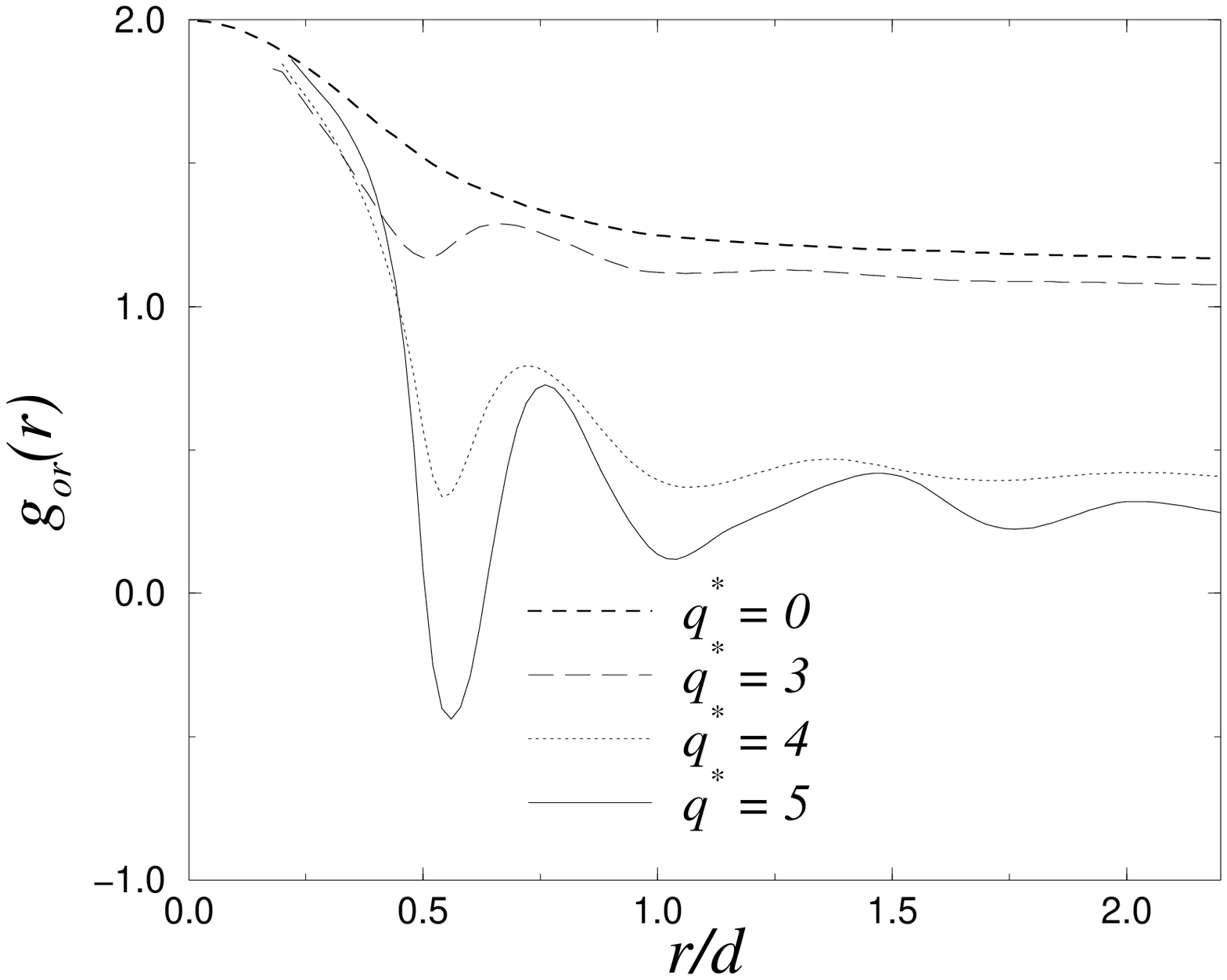,width=8.8cm,angle=0}
\caption{Effect of charge on the distributions $g(r)$ and $g_{or}(r)$, for
$\rho^*=5.0$ and a fixed screening length $\kappa d = 8$. }
\label{fig:jjwgrho5}
\end{figure}
\end{center}

Finally, we have performed exploratory runs at higher densities
where a nematic phase would be observed in the uncharged system.
The results of Fig. \ref{fig:jjwgrho5} for a high density 
$\rho^* = 5.0$ show a rich local structure both for the
$g(r)$ and the orientational $g_{or}(r)$, very different from
that observed at lower densities. The reference uncharged system 
($q^*=0$) with its strong nematic plateau is displayed for comparison. 
Inclusion of electrostatic interactions preserves the long range 
nematic order, but decreases its strength (the height of the plateau).
Such a nematic order was clearly absent at lower densities (see
Fig. \ref{fig:jjwkd3}). The peaks of $g(r)$ appear correlated to those
of $g_{or}(r)$, which points to the existence of oriented micro-domains 
with parallel platelets, and a higher nematic order than the mean one.
The complex behaviour of the distribution functions of Fig. \ref{fig:jjwgrho5}
calls for more thorough investigations at high densities.
In particular the validity of the large distance expansion 
(\ref{eq:dlvodiscs}) may be questionable at high $\rho^*$.

\section{Conclusion}
\label{sec:concl}

In this article we devise an electrostatic pair potential for both highly
anisotropic and highly charged objects (of vanishing internal volume)
based on a linearized Poisson-Boltzmann theory. On the example 
of infinitely thin platelets, we
show that the anisotropic shape of the object results in an anisotropic
Yukawa potential even at large distances.
The anisotropy is an increasing function of $\kappa r_0$ and becomes
very marked in the colloidal limit $\kappa r_0 >1$. 
To account for the non-linearities in the Poisson-Boltzmann theory,
when the bare charge of the platelets is large, we introduce an effective
or renormalized charge. This approximation yields an effective 
pair potential for highly charged platelets.

The addition of salt generically results in two antagonists effects,
irrespective of colloid geometry: the range
of the effective potential decreases, but its amplitude increases.
This interplay appears quite subtle, and able to discriminate between 
spherical and disc-like colloids by producing a re-entrant
melting line for discs. This phenomenon appears at first incompatible 
with naive DLVO expectation. Moreover, the corresponding threshold density
for discs is smaller than the isotropic/nematic coexistence for
uncharged plates, but still higher than its experimental 
``Fluid/Solid'' counterpart.

Preliminary results of Monte-Carlo simulations for particles interacting
through an anisotropic potential provide informations on the mesoscopic
structure of the dense phase. In particular, we show that the resulting
structure is very sensitive upon increasing the density. We have also
observed a ``re-entrant'' melting curve.
These effects are in 
qualitative agreement with the experimental phase diagram of Laponite
suspensions found in the literature, even if this topic is still 
under debate.

Our approach suffers from several weaknesses.  
a) the non-linearities in the Poisson-Boltzmann theory are accounted for at the
level of charge renormalization, 
b) in this framework, we expect 
the effective charge to be better approximated with an anzatz where 
the particle effectively behaves as a
constant electrostatic potential object \cite{Trizac2}, 
whereas we considered the platelets to be of constant surface charge,
c) we work at the level of pair
potential and correlation effects beyond the mean-field Poisson-Boltzmann
approach might play a role,
d) the precise role of van der Waals forces has not been assessed,
and remains obscure for clay platelets.
Further theoretical and numerical work is needed, 
but we hope that the arguments presented here provide a first hint 
into the full problem.
At least, our results strongly suggest that the combined effect of
anisotropy and charge condensation have a significant and non-trivial 
qualitative influence on the phase diagram of highly charged colloids.
These features could well be of prime importance in our understanding of
the thermodynamics of clay suspensions.

Acknowledgments. We thank J.-P. Hansen, H. Lekkerkerker, P. Levitz, A. Delville,
D. Bonn, T. Nicolai, B. Jancovici and F. van Wijland for fruitful discussions.

\appendix
\section{}
In this appendix, we show that the solution of Eq. (\ref{eq:lpb})
is not given by convolution product (\ref{eq:convol}) for polyions
having a non zero internal volume. For simplicity, we consider 
a charged hard sphere of radius $r_0$ 
(which consequently excludes micro-ions from its interior), with a 
uniform surface charge $\sigma$. The solution of Eq. (\ref{eq:lpb})
takes the well known DLVO form
\be
\phi(r) \, = \, Z \, \lb\,
\, \frac{e^{\kappa r_0}}{1+\kappa r_0}\, \frac{e^{-\kappa r}}{r},
\label{eq:appdlvo}
\ee
where $Ze = 4 \pi r_0^2 \sigma$ denotes the total bare charge. 
Alternatively, the convolution route of Eq. (\ref{eq:convol}) yields
\begin{eqnarray}
\phi_{\hbox{\scriptsize convol}}(r) &=& \lb \int_{\cal P} \frac{\sigma}{e} \,
\frac{\exp(-\kappa|{\bf r}-{\bf s}|)}{|{\bf r}-{\bf s}|} \,d^2{\bf s}
\label{eq:appconvol} \\
&=& Z \, \lb\,
\, \frac{\sinh(\kappa r_0)}{\kappa r_0}\, \frac{e^{-\kappa r}}{r} \quad
\hbox{for} \quad r \geq r_0.
\label{eq:appconvol2}
\end{eqnarray}
The origin of this discrepancy comes from the fact that upon summing up the
elementary surface contributions in (\ref{eq:appconvol}), the micro-ions
are allowed to enter the ``interior'' region $r \leq r_0$. The field
created is thus that of the shell with charge $Z$ (the colloid)
plus that of the ``plasma'' inside. The corresponding interior
charge $Z_{\hbox{\scriptsize in}}$ is easily computed from the expression 
of the electrostatic potential for $r\leq r_0$, which follows from a simple
permutation of $r$ and $r_0$ in expression (\ref{eq:appconvol2})
\be
\phi_{\hbox{\scriptsize convol}}^{\hbox{\scriptsize inside}}(r)\,=\, Z \, \lb\,
\, \frac{\sinh(\kappa r)}{\kappa r}\, \frac{e^{-\kappa r_0}}{r_0} .
\ee
We obtain
\begin{eqnarray}
Z_{\hbox{\scriptsize in}} &=& -\frac{\kappa^2}{4\pi\lb} \int_0^{r_0} 
\phi_{\hbox{\scriptsize convol}}^{\hbox{\scriptsize inside}}(r)\,d^3{\bf r} \\
&=& - Z e^{-\kappa r_0} \left[ \cosh(\kappa r_0) -
\frac{\sinh(\kappa r_0)}{\kappa r_0}\right].
\end{eqnarray}
Gathering results, a straightforward calculation allows to rewrite
Eq. (\ref{eq:appconvol2}) in the form
\be
\phi_{\hbox{\scriptsize convol}}(r) \,= \, Z \, \lb\,
\, \frac{\sinh(\kappa r_0)}{\kappa r_0}\, \frac{e^{-\kappa r}}{r} 
\,=\, \left( Z + Z_{\hbox{\scriptsize in}} \right)\, \lb\,
\frac{e^{\kappa r_0}}{1+\kappa r_0}\, \frac{e^{-\kappa r}}{r},
\ee
so that the DLVO structure of expression (\ref{eq:appdlvo})
is recovered. 

The argument given here bears some similarities with the proof of the equivalence
between two charged hard spheres models with a uniform background \cite{Hansennote}
(note that the equivalent of the background, the micro-ion density, is 
not uniform in the present situation). 
Finally, a related remark is that application of expression (\ref{eq:vabconvol})
for spheres does not give the DLVO pair potential (\ref{eq:dlvo})
{\em even if the correct one body potential (\ref{eq:appdlvo}) is used}
[use of expression (\ref{eq:appconvol}) is equally incorrect]. 
A simple way to recover (\ref{eq:dlvo}) is through the stress tensor
\cite{Belloni2}.




\newpage

\vskip 5mm

\vskip 1cm

\vskip 1cm

\begin{references}
\bibitem{Lapo}
Laponite particles may be considered as approximately monodisperse charged 
discs with radius 150$\,$\AA~ and thickness 10$\,$\AA.
\end{references}
\end{document}